\documentclass[10pt,letterpaper,twocolumn]{article} 

\usepackage{ol2}
\usepackage[draft]{hyperref}
\usepackage{amsmath}

\begin{document}

\twocolumn[ 

\title{Fractional Schr\"{o}dinger equation in optics}


\author{Stefano Longhi}

\address{Dipartimento di Fisica, Politecnico di Milano and Istituto di Fotonica e Nanotecnologie del Consiglio Nazionale delle Ricerche, Piazza L. da Vinci 32, I-20133 Milano, Italy (stefano.longhi@polimi.it)}

\begin{abstract}
In quantum mechanics, the space-fractional Schr\"{o}dinger equation provides a natural extension of the standard Schr\"{o}dinger equation when the Brownian trajectories in Feynman path integrals are replaced by Levy flights. Here an optical realization of the fractional Schr\"{o}dinger equation, based on transverse light dynamics in aspherical optical cavities, is proposed. As an example  a laser implementation of the fractional quantum harmonic oscillator is presented, in which dual Airy beams can be selectively generated under off-axis longitudinal pumping.
\end{abstract}


 ] 

\noindent
In the past few decades, several theoretical studies have speculated about possible extensions of standard quantum mechanics. While the implications and usefulness of such theories are still a matter of debate, some of their ideas and methods  have found interest beyond their original framework, with potential applications to science and technology. An example is provided by $\mathcal{PT}$-symmetric (non-Hermitian) extension of quantum mechanics \cite{B1}, which is attracting a growing interest in optics (see, for instance, \cite{B2,B3,B4,B5,B6} and references therein). Space-fractional quantum mechanics (SFQM), introduced in 2000
by Laskin \cite{L1,L2,L3}, provides a different generalization of the standard quantum mechanics that arises when the Brownian trajectories in Feynman path integrals are replaced by Levy flights. SFQM provides an interesting fractional physical model in quantum physics \cite{libro}. Most of current studies on SFQM have focused on the mathematical aspects of the theory \cite{M1,M2,M3,M4,M5} and have highlighted subtle issues that arise from  the non-local nature of the kinetic operator \cite{M3}, however very few studies have considered possible physical realizations or applications of SFQM \cite{libro,E1}.  In Ref. \cite{E1} a condensed-matter realization of SFQM based on Levy crystals was suggested, however its experimental implementation seems very challenging owing to the need to specially tailoring non-nearest neighbor  hopping in the lattice.\par
In this Letter we suggest a different route toward a physical implementation of SFQM, which is based on transverse light dynamics in aspherical optical resonators. As an example, we discuss the design of an optical resonator that realizes 
the fractional quantum harmonic oscillator \cite{L1,L3} (also known as massless relativistic quantum oscillator \cite{O1,O2}). Interestingly, we show that dual Airy beams \cite{dual} can be generated from such an optical cavity.\par
The space-fractional Schr\"{o}dinger equation (SFSE) was derived by Laskin is a series of papers using a Feynman path integral approach over Levy trajectories \cite{L1,L2,L3}. In the one spatial dimension and assuming $\hbar=1$ it reads
\begin{equation}
i \frac{\partial \psi}{\partial t} = \mathcal{D}_{\alpha} \left( - \frac{\partial^2}{\partial x^2} \right)^{\alpha /2} \psi + V(x)  \psi
\end{equation}
where $\alpha$ is the Levy index ($1 <  \alpha \leq 2$), $\mathcal{D}_{\alpha}$ is a scale constant, $V(x)$ is the external potential, and $\psi=\psi(x,t)$ is the particle wave function. In Eq.(1) the kinetic term in the Hamiltonian is represented by the quantum Riesz derivative (fractional Laplacian) of order $\alpha$, which is defined by
\begin{equation}
\left( -\frac{\partial^2 }{\partial x^2} \right)^{\alpha / 2} \psi(x)=\frac{1}{2 \pi} \iint_{- \infty}^{\infty} dp d\xi  |p|^{\alpha} \psi( \xi) \exp[ip(x-\xi)].
\end{equation}
The ordinary Schr\"{o}dinger equation is obtained in the limiting case $\alpha=2$. In Ref.\cite{E1}, a realization of the SFQM based on lattice dynamics in so-called Levy crystals was suggested. Here we propose an optical realization of the SFSE based on transverse light dynamics in an optical cavity, in which we exploit the properties of Fourier optics to realize the fractional Laplacian \cite{Opt1,Opt2}.  We consider an effective one-dimensional optical resonator problem with transverse $x$ spatial direction. Such an analysis holds for e.g. an optical cavity with astigmatic optical elements in which separation of variables in the two transverse spatial dimensions $x$ and $y$ can be applied and the lowest-order transverse mode in the $y$ direction is excited. A schematic of the effective one-dimensional Fabry-Perot optical resonator is depicted in Fig.1(a). It comprises two flat end mirrors and two converging lens of focal length $f$ in a typical $4f$ configuration, with two thin phase masks with transmission functions $t_1(x)= \exp[-if(x)/2]$ and $t_2(x)=\exp[-iV(x)/2]$, where
\begin{equation}
f(x)=\beta |x|^{\alpha}.
\end{equation} 
\begin{figure}[htb]
\centerline{\includegraphics[width=8.2cm]{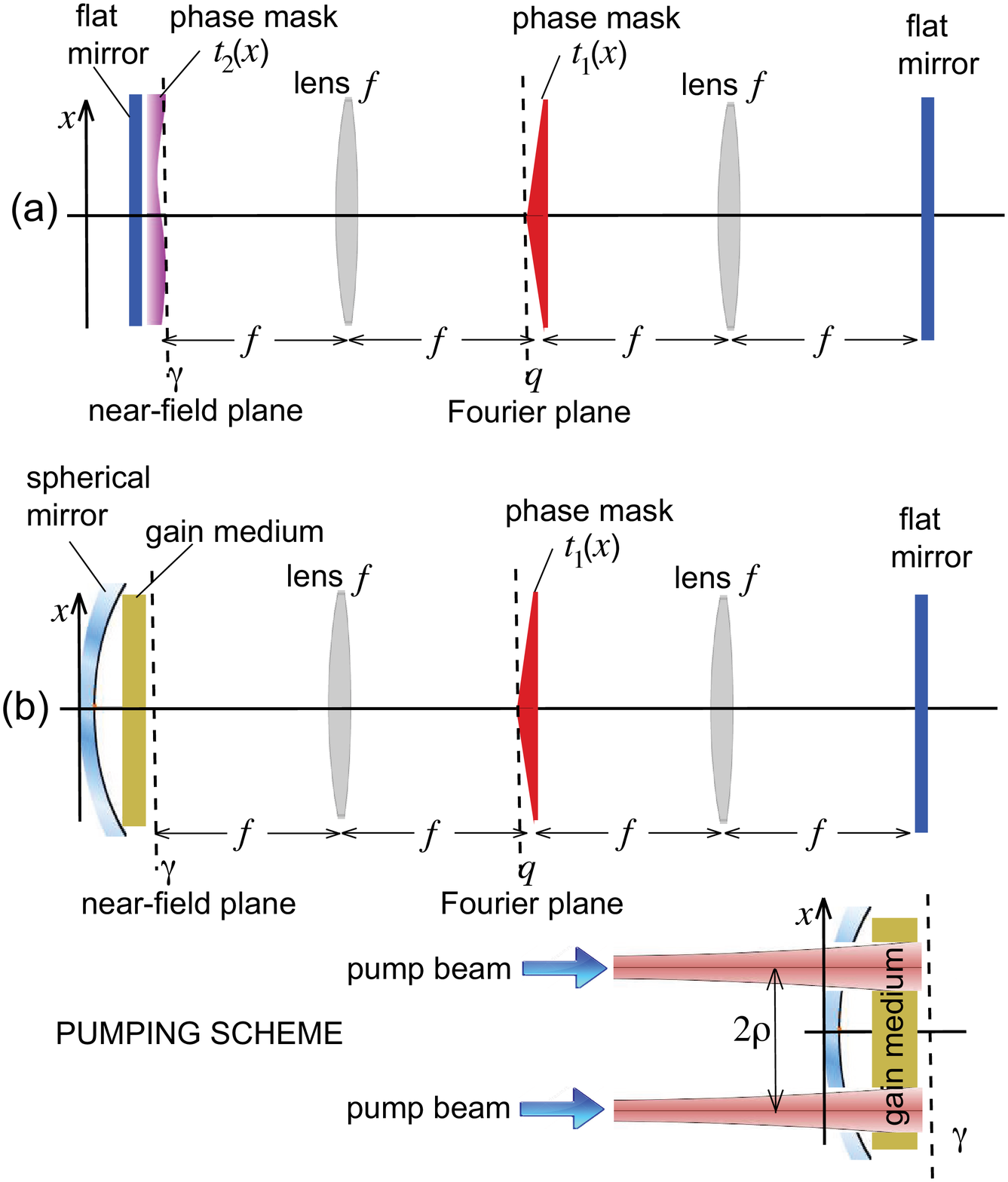}} \caption{ \small
(Color online) (a) Schematic of an optical resonator in a 4-$f$ configuration that realizes the SFSE. (b) Schematic of the optical resonator that realizes the fractional quantum harmonic oscillator. The phase mask $t_2(x)$ and flat end mirror in (a) are replaced by a spherical mirror. The pumping scheme to selectively excite higher-order dual-Airy TEM$_{n}$ modes is shown in the lower panel.}
\end{figure}
In the scalar and paraxial approximations, wave propagation at successive transits inside the optical cavity can be readily obtained from the generalized Huygens integral by standard methods \cite{Siegman}. Neglecting finite aperture effects and indicating by $F^{(n)}(x)=i^{n} \psi^{(n)}(x)$ the field envelope of the progressive wave at the reference plane $\gamma$ in the cavity and at the $n$-th round-trip, in the absence of gain/loss elements one can write 
\begin{equation}
\psi^{(n+1)}(x)= \int d \theta  K(x, \theta) \psi^{(n)}(\theta),
\end{equation}
 where the kernel  $K$ of the integral transformation is given by  
\begin{eqnarray}
K(x, \theta)   & = &  \left( \frac{1}{\lambda f} \right) \times \\
 & \times &  \int d \xi  \exp \left[ -if(\xi)-iV(x) +\frac{2 \pi i \xi(x-\theta)}{\lambda f} \right] \;\;\;\;\;\; \nonumber 
\end{eqnarray}
and $\lambda= 2 \pi /k= c / \nu$ is the optical wavelength. 
The transverse modes TEM$_n$ and corresponding resonance frequencies of the resonator are obtained from the eigenfunctions $\psi_n(x)$ and corresponding eigenvalues  $\sigma_n$ of the Fredholm equation $\sigma_n \psi_n(x)=\int d \theta K(x,\theta) \psi_n(\theta)$ \cite{Siegman}. Note that, in the absence of the phase masks, i.e. for $V(x)=f(x)=0$, the $4f$ resonator is a self-imaging cavity, i.e. $K(x,\theta)=\delta(x-\theta)$, so that any field distribution is self-reproduced at each round trip. Here we are interested in the limiting case $V(x), f(x) \rightarrow 0$,  so that the field is nearly self-imaged and undergoes slow changes at each round trip. If the exponential term $\exp[-iV(x)-if(\xi)]$ entering in the kernel $K$ is expanded as  $\exp[-iV(x)-if(\xi)] \simeq 1-iV(x)-if(\xi)$, using Eqs.(2,3) one readily obtains
 \begin{equation}
 \psi^{(n+1)}(x) \simeq \left[1 -iV(x) \psi^{(n)}(x)-i \mathcal{D}_{\alpha} \left( -\frac{\partial^2}{\partial x^2} \right)^{\alpha/2} \right] \psi^{(n)}(x)
 \end{equation}
where we have set $\mathcal{D}_{\alpha} \equiv \beta (\lambda f / 2 \pi)^{\alpha}$. If we introduce the time variable $t$, normalized to the round-trip transit time $T_R$ of photons in the cavity, after setting $\psi(x,t)=\psi^{(n=t)}(x)$, $(\partial \psi / \partial t) \simeq \psi^{(n+1)}(x)-\psi^{(n)}(x)$, the temporal evolution of the transverse field envelope $\psi(x,t)$ at plane $\gamma$ is thus governed by the SFSE
\begin{equation}
i \frac{\partial \psi} {\partial t}= \left[ \mathcal{D}_{\alpha} \left( -\frac{\partial^2}{\partial x^2} \right)^{\alpha/2} +V(x)  \right] \psi(x).
\end{equation}
Note that at the Fourier plane $q$ the Fourier transform field 
\begin{equation}
\phi(x,t)=\sqrt{\frac{i}{\lambda f }} \int_{-\infty}^{\infty} d \xi \psi( \xi,t) \exp[2 i \pi x \xi / (\lambda f)]  
\end{equation}
of $\psi(x,t)$ is observed, which satisfies the SFSE in 'momentum space' \cite{M1}
\begin{eqnarray}
i \frac{\partial \phi}{\partial t} & = & \beta|x|^{\alpha} \phi + \\
& + & \frac{1}{\lambda f} \iint_{-\infty}^{\infty} d \xi d \theta V(\xi) \phi(\theta,t) \exp \left[ \frac{2 \pi i \xi(x-\theta)} {\lambda f} \right]. \nonumber
\end{eqnarray}
Therefore the transverse modes and resonance frequencies of the resonator of Fig.1(a) correspond to the eigenfunctions and corresponding energies of the SFSE with Levy index $\alpha$ in the external potential $V(x)$.\par As an example, let us consider the optical realization of the {\it fractional quantum harmonic oscillator} \cite{L1,L3}, which corresponds to a parabolic external potential $V(x) \propto x^2$. This case can be readily implemented by substitution of the phase mask and flat mirror at plane $\gamma$ by a simple spherical mirror of radius of curvature $R$ [see Fig.1(b)], corresponding to the potential 
\begin{equation}
V(x)=\frac{2 \pi} {\lambda R}x^2.
\end{equation}
The transverse modes (eigenfunctions) and corresponding frequencies of the resonator can be at best determined from the SFSE eigenvalue equation in momentum space [Eq.(9)], i.e. at the Fourier plane $q$. In fact, after setting $\phi(x,t)=\phi(x) \exp(-iEt)$, from Eqs.(9) and (10) one obtains the eigenvalue equation
\begin{equation}
E \phi(x)=\beta|x|^{\alpha} \phi(x)-\left( \frac{\lambda f^2}{2 \pi R} \right) \frac{d^2 \phi}{d x^2}.
\end{equation}
 A particularly interesting case which admits of an analytical solution is the one corresponding to $\alpha=1$, which is realized by using a one-dimensional refractive axicon in the Fourier plane $q$ . This limiting case of fractional quantum oscillator realizes the so-called massless relativistic harmonic oscillator, which is studied e.g. in Refs.\cite{O1,O2}. For $\alpha=1$ Eq.(11) reduces to the Airy equation, with eigenfunctions (transverse modes TEM$_n$) and corresponding energies $E_n$ given by
 \begin{eqnarray}
 \phi_n(x) & = &  \left( \frac{x}{|x|} \right)^n {\rm Ai} \left( \kappa |x|+r_n \right) \\\
 E_n & = & - \frac{\lambda f^2}{2 \pi R} \kappa^2 r_n
 \end{eqnarray}
 where ${\rm Ai}(x)$ is the Airy function, $\kappa \equiv (2 \pi \beta R/ \lambda f^2)^{(1/3)}$, and where $r_n$ are the zeros of $(d {\rm Ai} /dx)$ (for $n$ even) or of ${\rm Ai}(x)$ (for $n$ odd). Note that, since $r_n<0$, one has $E_n>0$. Interestingly, for large enough mode number $n$ the transverse modes TEM$_n$, as described by Eq.(12), mainly correspond to two displaced Airy beams that accelerate into opposite directions, the so-called dual Airy beams \cite{dual}. In an optical experiment, such higher-order modes can be selectively excited in a laser device under off-axial longitudinal pumping \cite{Longhi}. To this aim, let us introduce a thin gain medium close to the spherical mirror, see Fig.1(b). We assume that the medium is longitudinally-pumped by two off-axis Gaussian beams of spot size $w_p$ and transversely displaced by $2 \rho$. Taking into account the presence of the gain medium and the output coupling of the cavity, the round-trip propagation of the intracavity field $\psi^{(n)}(x)$ at plane $\gamma$ is modified as follows [compare with Eq.(4)]
 \begin{equation}
 \psi^{(n+1)}(x)=\sqrt{T} \exp[ g(x)] \int d \theta K(x,\theta) \psi^{(n)} (\theta)
 \end{equation}
where $T$ is the transmittance of the output coupler, $g(x)$ is the transversely-dependent round-trip gain in the active medium, and the kernel $K$ is given by Eq.(5). The gain parameter $g(x)$ is given by $g(x)= 2 \sigma  N(x) l$, where $l$ is the thickness of the gain medium, $\sigma$ is the stimulated cross section of the lasing transition and $N(x)$ is the population inversion in the medium. For the off-axis longitudinal pumping scheme shown in Fig.1(b) and neglecting saturation of the gain, we may assume 
\begin{equation}
g(x)=g_0 \exp \left[-2\frac{(x-\rho)^2}{w_p^2} \right]+g_0 \exp \left[-2\frac{(x+\rho)^2}{w_p^2} \right],
\end{equation}
 where $w_p$ is the pump beam spot size and $g_0$ is the peak gain parameter. The lowest-thresold lasing mode and corresponding laser threshold $g_0=g_{0th}$ can be numerically computed
from Eq.(14) by a standard Fox-Li iterative method \cite{Fox}. As the displacement $2 \rho$ between the two pumping beams is increased, higher-order TEM$_n$ dual-Airy transverse modes of the resonator with increasing mode number $n$ can be selectively excited. This is clearly shown in Fig.2. The figure shows the numerically-computed lowest-threshold transverse modes, both in the near-field plane $\gamma$ (left panels) and in the Fourier plane $q$ (right panels), as obtained by the Fox-Li iterative method for parameter values $f=1$ cm, $\lambda=1064$ nm (Nd:YAG laser), $T=97 \%$, $R= 50$ cm, $\beta=5 \times 10^{-3} \; \mu{\rm m}^{-1}$, $w_p=\sqrt{2} \times 20 \; \mu$m and for increasing values of $\rho$. For $\rho=0$, i.e. for on-axis pumping [Fig.2(a)], the lasing mode is the lowest-order transverse mode TEM$_0$, corresponding to $r_0 \simeq -1.05$ in Eq.(12). At $\rho= 170 \; \mu$m [Fig.2(b)], the lasing mode is the transverse mode TEM$_{6}$ with mode index $n=6$, corresponding to $r_6 \simeq -6.16$. Finally, at $\rho= 256  \; \mu$m [Fig.2(c)] the lowest-threshold mode is the transverse mode TEM$_{16}$ with mode index $n=16$, corresponding to $r_{16} \simeq -11.48$.  The threshold values of lasing in three cases are $g_{0th} \simeq 0.028$ in Fig.2(a), $g_{0th} \simeq 0.075$ in Fig.2(b) and $g_{0th} \simeq 0.098$ in Fig.2(c). Therefore the laser system schematically shown in Fig.1(b) provides an optical realization of the fractional quantum harmonic oscillator, in which the different modes of the oscillator can be selectively excited by suitable off-axis pumping. For large off-axis pumping, the lasing modes correspond to dual Airy beams, and thus our system provides an interesting example (additional to  those previously reported in  Refs.\cite{Longhi,Arie}) where Airy-type beams can be directly generated from a laser oscillator. It should be finally noted that the above analysis does not consider mode confinement in the other transverse $y$ direction, however such a confinement can be readily obtained using an astigmatic optical cavity. Namely, let us consider the optical cavity of Fig.1(b) in which the two converging lenses (as well as the phase mask at the Fourier plane $q$) are astigmatic and acts along the spatial $x$ direction solely, whereas the end spherical mirror is circularly symmetric and so it acts in both $x$ and $y$ directions. In this case the resonator TEM$_{n,m}$ transverse modes can be readily obtained by separation of variables, i.e. they factorize as the products of the modes of two effective one-dimension resonators in the $x$ and $y$ directions. For example, at the Fourier plane $q$ the transverse modes TEM$_{n,m}$ are given by $\phi_{n,m}(x,y)=\phi_n(x) Y_m(y)$, where $\phi_n(x)$ are the dual Airy modes given by Eq.(12) whereas 
 \begin{equation}
 Y_m(y)=H_m \left( \frac{\sqrt{2}y}{w_y}  \right) \exp \left( -\frac{y^2}{w_y^2} -i \frac{\pi y^2}{\lambda R_y} \right)
\end{equation}
are standard Gauss-Hermite modes of index $m$ with beam spot size $w_y$ and radius of curvature $R_y$. The values of $w_y$ and $R_y$ at the Fourier plane $q$ can be readily computed by standard Gaussian beam analysis of the resonator in the $y $direction and read 
\begin{eqnarray}
w_y & = & \sqrt{\frac{\lambda f}{\pi}} \left( \frac{R}{4f} -1 \right)^{1/4} \sqrt{\frac{4R-12f}{R-4f}} \\
R_y & = & 2R-6f.
\end{eqnarray}
For a two-beam longitudinal pumping with $2 \rho$ off-axis displacement in the $x$ direction solely, i.e. for a two-dimensional optical gain distribution
$g(x,y)=g_0 \exp [-2(x-\rho)^2/w_p^2-2y^2/w_p^2]+$ $+g_0 \exp [-2(x+\rho)^2/w_p^2-2y^2/w_p^2]$, the lowest order transverse mode $m=0$ in the $y$ direction is excited. In this case, as $\rho$ is increased from zero, higher order dual-Airy transverse modes TEM$_{n,0}$ in the $x$ direction are excited, and the resulting two-dimensional intensity distributions of lasing modes are those depicted in Fig.3.

\begin{figure}[htb]
\centerline{\includegraphics[width=8.2cm]{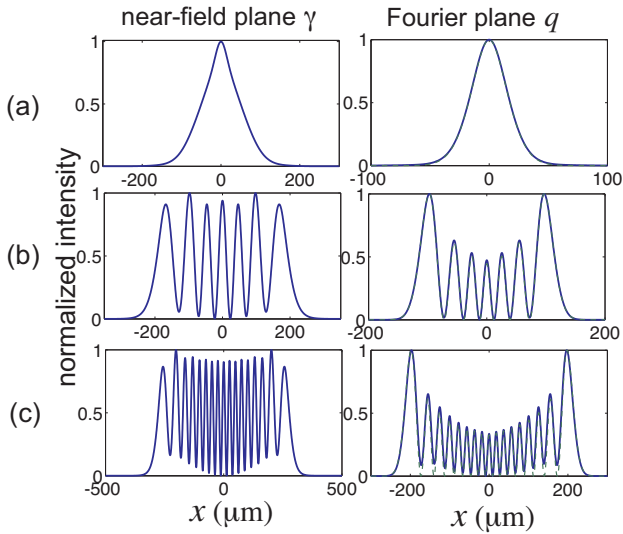}} \caption{ \small
(Color online) Numerically-computed intensity distribution of the lowest-threshold lasing modes in the optical cavity of Fig.1(b) for increasing values of the off-axis pump displacement $\rho$. The left panels show to the transverse intensity distributions in the near-field plane $\gamma$, whereas the right panels are the corresponding intensity distributions in the Fourier plane $q$. (a) $\rho=0$ (on-axis pumping), (b) $\rho=170 \; \mu$m, and (c) $\rho=256 \; \mu$m. The other parameter values are given in the text. The dashed curves in the right panels, almost overlapped with the solid ones, show the intensity distributions of the transverse modes TEM$_n$, as predicted by Eq.(12), for $n=0$ [in (a)], $n=6$ [in (b)], and $n=16$ [in (c)].}
\end{figure}

\begin{figure}[htb]
\centerline{\includegraphics[width=8.3cm]{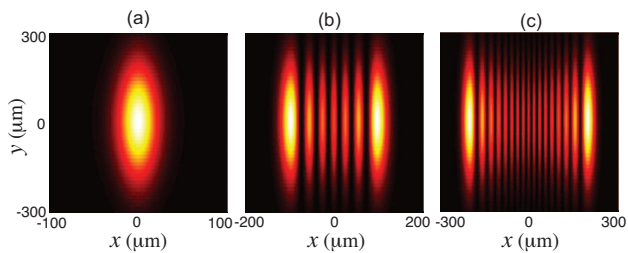}} \caption{ \small
(Color online) Two-dimensional intensity distributions of the lasing modes at the Fourier plane $q$ in the astigmatic cavity of Fig.1(b), taking into account confinement in the other spatial direction $y$ provided by the spherical end mirror. For increasing values of the the pump displacement $\rho$ higher-order TEM$_{n0}$ transverse modes are excited: (a) $\rho=0$ (on-axis pumping), (b) $\rho=170 \; \mu$m, and (c) $\rho=256 \; \mu$m.}
\end{figure}

\par 
In conclusion, an optical implementation of the fractional Schr\"{o}dinger equation has been proposed, which is based on transverse laser dynamics in aspherical optical resonators. As an example, a laser design for the generation of dual-Airy states of a fractional quantum harmonic oscillator has been presented. The ability in optics to implement fractional pseudo-derivative operators could be further exploited to simulate other quantum models involving fractional operators, such as the relativistic Salpeter wave equation \cite{Sal1,Sal2}. Our results indicate that optics can provide a laboratory tool where fractional models developed in quantum physics can become experimentally accessible. Fractional models could also provide new interesting ways to mold diffraction of light and to generate novel beam solutions \cite{Vega}.

\newpage

\footnotesize {\bf References with full titles}\\
\\

\noindent
1. C.M. Bender, {\it Making sense of non-Hermitian Hamiltonians}, Rep. Prog. Phys. {\bf 70}, 947 (2007).\\
2. R. El-Ganainy, K. G. Makris, D. N. Christodoulides, and Z.H. Musslimani, {\it Theory of coupled optical $\mathcal{PT}$ symmetric structures}, Opt. Lett. {\bf 32}, 2632 (2007).\\
3. K. G. Makris, R. El-Ganainy, D. N. Christodoulides, and Z. H. Musslimani, {\it Beam Dynamics in $\mathcal{PT}$-Symmetric Optical Lattices}, Phys. Rev. Lett. {\bf 100}, 103904 (2008).\\
4. C. E. R\"{u}ter, K.G. Makris, R. El-Ganainy, D.N. Christodoulides, M. Segev, and D. Kip, {\it Observation of parityÐtime symmetry in optics}, Nature Phys. {\bf 6}, 192 (2010).\\
5. S. Longhi, {\it $\mathcal{PT}$ symmetric laser-absorber}, Phys. Rev. A {\bf 82}, 031801 (2010).\\
6. A. Regensburger, C. Bersch, M.-A. Miri, G. Onishchukov, D.N. Christodoulides, and	U. Peschel, {\it Parity-time synthetic photonic lattices}, Nature {\bf 488}, 167 (2012).\\
7. N. Laskin, {\it Fractional Quantum Mechanics and Levy Path Integrals}, Phys. Lett. A {\bf 268}, 298 (2000).\\
8. N. Laskin, {\it Fractional Quantum Mechanics}, Phys. Rev. E {\bf 62}, 3135 (2000).\\
9. N. Laskin, {\it Fractional Schr\"{o}dinger equation}, Phys. Rev. E {\bf 66}, 056108 (2002).\\
10. R. Herrmann, {\it Fractional Calculus: An Introduction for Physicists} (World Scientific,  2011), chaps. 9 and 10. 
11. J. Dong and M. Xu, {\it Some solutions to the space fractional Schr\"{o}dinger equation using momentum representation method}, J. Math. Phys. {\bf 48}, 072105  (2007).\\
12. E. Capelas de Oliveira and J. Vaz, {\it Tunneling in fractional quantum mechanics}, J. Phys. A {\bf 44},185303 (2011).\\ 
13. Y. Luchko, {\it Fractional Schr\"{o}dinger equation for a particle moving in a potential well}, J. Math. Phys. {\bf 54}, 012111 (2013).\\
14. M. Zaba and P. Garbaczewski, {\it Solving fractional Schr\"{o}dinger-type spectral problems: Cauchy oscillator and Cauchy well}, J. Math. Phys. {\bf 55}, 092103 (2014).\\
15. J.D. Tare, J. Perico, and H. Esguerra, {\it Transmission through locally periodic potentials in space-fractional quantum mechanics}, Physica A {\bf 407}, 43 (2014).\\
16. B. A. Stickler, {\it Potential condensed-matter realization of space-fractional quantum mechanics: The one-dimensional Levy crystal}, Phys. Rev. E {\bf  88}, 012120 (2013).\\
17. K. Kowalski and J. Rembielinski, {\it Relativistic massless harmonic oscillator}, Phys. Rev. A {\bf 81}, 012118 (2010).\\
18. J. L\"{o}rinczi and J. Malecki, "Spectral properties of the massless relativistic harmonic oscillator", J. Diff. Eq. {\bf 253},  2846 (2012).\\ 
19. C.-Y. Hwang, D. Choi, K.-Y. Kim, and B. Lee, {\it Dual Airy beams}, Opt. Expr. {\bf 18}, 23504 (2010).\\
20. H. Kasprzak, {\it Differentiation of a noninteger order and its optical implementation}, Appl. Opt. {\bf 21}, 3287 (1982).\\
21. J.A. Davis, D.A. Smith, D.E. McNamara, D.M. Cottrell, and J. Campos, {\it Fractional derivatives-analysis and experimental
implementation}, Appl. Opt. {\bf 40}, 5943 (2001).\\
22. A.E. Siegman, {\it Lasers} (University Science, Mill Valley, Calif., 1986), Chaps. 15, 16 and 20.\\
23. S. Longhi, {\it Airy beams from a microchip laser}, Opt. Lett. {\bf 36}, 716 (2011).\\
24. A. O. Fox and T. Li, {\it Resonant modes in a maser interferometer}, Bell Syst. Tech. J. {\bf 40}, 453 (1961).\\
25. G. Porat, I. Dolev, O. Barlev, and A. Arie, {\it Airy beam laser}, Opt. Lett. {\bf 36}, 4119 (2011).\\
26. E.E. Salpeter, {\it Mass Corrections to the Fine Structure of Hydrogen-Like Atoms}, Phys. Rev. {\bf 87}, 328 (1952).\\
27. P. Cea, P. Colangelo, G. Nardulli, G. Paiano, and G. Preparata, {\it WKB approach to the Schr\"{o}dinger equation with relativistic kinematics}, Phys. Rev. D {\bf 26},  1157 (1082).\\
28. J.C. Gutierrez-Vega, {\it Fractionalization of optical beams: I. Planar analysis}, Opt. Lett. {\bf 32}, 1521 (2007).

\end{document}